\documentclass[a4paper,11pt]{article}
\usepackage{jcappub}

\usepackage{amsthm}
\usepackage{graphicx}
\usepackage{xcolor}
\usepackage{subcaption}
\usepackage{physics}
\usepackage{braket}
\usepackage{amsmath}
\usepackage{amssymb}
\usepackage{enumerate}
\usepackage{mathtools}
\usepackage{etoolbox}
\usepackage{algorithm}
\usepackage{algpseudocode}
\usepackage{bbm}
\usepackage{here}
\usepackage{comment}
\usepackage{bm}
\usepackage{acronym}
\usepackage{siunitx}
\usepackage{ragged2e}
\usepackage{booktabs}

\newcounter{algo}

\newcommand{\beae}[1]{\begin{equation}\begin{aligned} #1 \end{aligned}\end{equation}}

\newcommand{\bae}[1]{\begin{align} #1 \end{align}}
\newcommand{\bce}[1]{\begin{cases} #1 \end{cases}}

\newcommand{\bme}[1]{\begin{multline} #1 \end{multline}}
\newcommand{\bmte}[1]{\begin{multlined}[t] #1 \end{multlined}}

\newcommand{\bk}{\mathrm{bk}}

\newcommand{\Mpl}{M_\mathrm{Pl}}

\newcommand{\ini}{\mathrm{ini}}
\newcommand{\rmL}{\mathrm{L}}
\newcommand{\rmS}{\mathrm{S}}

\newcommand{\NL}{N_\rmL}
\newcommand{\NS}{N_\rmS}

\newcommand{\uc}{\mathrm{c}}
\newcommand{\sfH}{\mathsf{H}}
\newcommand{\bfk}{\mathbf{k}}
\newcommand{\calN}{\mathcal{N}}
\newcommand{\calP}{\mathcal{P}}
\newcommand{\calB}{\mathcal{B}}

\newcommand{\bfx}{\mathbf{x}}
\newcommand{\bfy}{\mathbf{y}}
\newcommand{\bfX}{\mathbf{X}}
\newcommand{\bfphi}{\boldsymbol{\phi}}
\newcommand{\bfpi}{\boldsymbol{\pi}}

\newcommand{\delN}{\delta N}
\newcommand{\delcalN}{\delta \calN}
\newcommand{\bfcalX}{\boldsymbol{\mathcal{X}}}

\newcommand{\bfXS}{\mathbf{X}_\rmS}

\abstract{
The bispectrum of cosmological perturbations is an important probe of inflation and the underlying particle physics.
However, calculating it becomes challenging in inflation models where inflaton dynamics is dominated by quantum diffusion rather than slow-roll, especially in the multi-field case.
In this paper, employing the stochastic-$\delta \calN$ formalism, we propose a Monte Carlo-based method to calculate the bispectrum in the squeezed limit, as an extension of the method for the power spectrum that we previously proposed.
Our method involves generating paths of inflatons' time evolution that branch only several times, avoiding nested path generation, which incurs a prohibitive computational cost.
As numerical demonstrations, we apply the proposed method to single- and double-field chaotic inflation for validation, and to hybrid inflation with mild waterfall, for which, to the best of our knowledge, the bispectrum is calculated for the first time in this work.
}

\begin{document}

\title{\boldmath Calculating the squeezed bispectrum in stochastic inflation by Monte Carlo simulation
}

\author[a]{Koichi Miyamoto}
\author[b]{and Yuichiro Tada}

\affiliation[a]{Center for Quantum Information and Quantum Biology, The University of Osaka, 1-2 Machikaneyama, Toyonaka, Osaka, 560-0043, Japan}
\affiliation[b]{Department of Applied Physics, University of Fukui, \\
Bunkyo 3-9-1, Fukui 910-8507, Japan}

\emailAdd{miyamoto.kouichi.qiqb@osaka-u.ac.jp}
\emailAdd{ytada@u-fukui.ac.jp}

\date{\today}

\maketitle

\acrodef{PDF}{probability density function}
\acrodef{SDE}{stochastic differential equation}
\acrodef{EOI}{end of inflation}
\acrodef{EM}{Euler--Maruyama}
\acrodef{PRNG}{pseudo-random number generator}
\acrodef{CR}{consistency relation}
\acrodef{CRN}{common random numbers}
\acrodef{PBH}{primordial black hole}

\section{Introduction}

Analysing cosmological perturbations generated in various inflation models is a central issue in inflationary cosmology, since features of these perturbations hold the key to understanding the history of the early universe and the underlying particle physics. 
For this purpose, various approaches have been proposed.
On the one hand, the $\delta N$ formalism~\cite{Starobinsky:1985ibc,Salopek:1990jq,Sasaki:1995aw,Wands:2000dp,Lyth:2004gb}, which relates fluctuations of the e-fold number $N$ during inflation in different spatial patches to primordial perturbations, is a widely-used one.
On the other hand, a probabilistic formalism called stochastic inflation (see Refs.~\cite{Starobinsky:1982ee,Starobinsky:1986fx,Nambu:1987ef,Nambu:1988je,Kandrup:1988sc,Nakao:1988yi,Nambu:1989uf,Mollerach:1990zf,Salopek:1990re,Linde:1993xx,Starobinsky:1994bd} for the first works and also Ref.~\cite{Cruces:2022imf} for a recent review) has been proposed: in the formalism, we consider the inflaton fields $\boldsymbol{\varphi}$ and their conjugate momenta $\boldsymbol{\varpi}$ coarse-grained on a superHubble scale and describes their stochastic time-evolution driven by quantum fluctuations using the Langevin equation.
Their combination, the \emph{stochastic-$\delta \calN$ formalism}~\cite{Fujita:2013cna,Fujita:2014tja,Vennin:2015hra,Ando:2020fjm,Animali:2024jiz}, is often useful, especially in diffusion-dominated cases, where inflatons traverse a very flat region in their potential, and thus their stochastic movement dominates over the usual slow-roll dynamics.
Such a scenario has attracted attention because primordial curvature perturbations $\zeta$ of corresponding scale can be amplified, which leads to interesting cosmological phenomena such as the production of \acp{PBH}, a candidate for dark matter.
In fact, many papers~\cite{Kawasaki:2015ppx,Assadullahi:2016gkk,Vennin:2016wnk,Pattison:2017mbe,Ezquiaga:2018gbw,Noorbala:2018zlv,Firouzjahi:2018vet,Noorbala:2019kdd,Kitajima:2019ibn,Prokopec:2019srf,Ezquiaga:2019ftu,Firouzjahi:2020jrj,De:2020hdo,Figueroa:2020jkf,Pattison:2021oen,Figueroa:2021zah,Tada:2021zzj,Ezquiaga:2022qpw,Ahmadi:2022lsm,Nassiri-Rad:2022azj,Animali:2022otk,Tomberg:2022mkt,Gow:2022jfb,Rigopoulos:2022gso,Briaud:2023eae,Asadi:2023flu,Tomberg:2023kli,Tada:2023fvd,Tokeshi:2023swe,Raatikainen:2023bzk,Miyamoto:2024hin,Tokeshi:2024kuv,Kuroda:2025coa,Takahashi:2025hqt} have conducted stochastic inflation-based analyses in aforementioned scenarios both analytically and numerically, aiming to calculate quantities of cosmological interest, such as the power spectrum $\mathcal{P}_\zeta$ of primordial perturbations.

Nevertheless, conducting such analyses in general cases, especially multi-field models, is challenging, whereas there are well-motivated multi-field diffusion-dominated scenarios such as mild waterfall transition in hybrid inflation~\cite{Fujita:2014tja,Kawasaki:2015ppx,Tada:2023pue,Tada:2023fvd,Tada:2024ckk,Murata:2025onc,Murata:2026yqb}.
In such a setting, we basically resort to numerical methods, such as Monte Carlo simulation \cite{glasserman2004monte}, a widely-used approach for analysing multi-dimensional random processes.
In fact, Ref.~\cite{Fujita:2014tja} proposed a Monte Carlo-based method, where we generate many paths of time evolution of $(\boldsymbol{\varphi},\boldsymbol{\varpi})$ and use them to calculate quantities such as $\mathcal{P}_\zeta$.
However, this method still suffers from high computational cost, whether the inflaton is single or multiple, because it requires a nested Monte Carlo simulation: we generate many paths, and then, from selected points on each path, we further generate many branch paths.

Recently, the authors of this paper proposed a more efficient Monte Carlo-based method for calculating $\mathcal{P}_\zeta$ that avoids nested path generation \cite{miyamoto2025calculating}.
Building upon the formula of $\mathcal{P}_\zeta$ given by Ando and Vennin (AV) \cite{Ando:2020fjm}, Ref. \cite{miyamoto2025calculating} found an estimator of the statistic involved in AV's formula, which we can calculate by generating not many but only two branches per branching point.
Then, Ref. \cite{miyamoto2025calculating} conducted some numerical demonstrations for some inflation models, including the hybrid model with mild waterfall transition, in which we obtained $\mathcal{P}_\zeta$ with a peak at the scale corresponding to the transition.

In this paper, our next goal is to calculate the bispectrum $\calB_\zeta$ of curvature perturbations, especially in the squeezed limit.
It has attracted large attention as a reflection of the non-Gaussianity of cosmological perturbations.
Although the bispectrum at large scales has been tightly constrained by cosmic microwave background observations, with results consistent with zero \cite{akrami2020planck}, there is still a possibility of a sizable bispectrum at smaller scales, which may lead to observable cosmological consequences.
In particular, in the aforementioned scenarios where $\mathcal{P}_\zeta$ is amplified at a specific scale, $\calB_\zeta$ may be amplified as well.
To investigate such a case and others, we propose a Monte Carlo-based method for calculating the squeezed bispectrum.
We build upon the formula by Ref. \cite{Tada_2017}, which gives the squeezed bispectrum by the correlation between the long-scale perturbation and the small-scale power spectrum, differentiated by the scale.
Rewriting the formula, we derive a Monte Carlo estimate of the squeezed bispectrum, which we can calculate via path generation with only several branchings, thereby avoiding a nested Monte Carlo simulation.
As a third-order quantity of perturbations, it is more difficult to calculate the bispectrum than the power spectrum, a second-order quantity, especially by the Monte Carlo method, which inevitably accompanies a statistical error.
Nevertheless, employing the technique called the \ac{CRN} method~\cite{glasserman2004monte}, which is often used for calculating the derivative of a function estimated by the Monte Carlo method, we make an effort to suppress the error in the estimated bispectrum.

We then conduct numerical demonstrations of the proposed method for three inflation models: single-field chaotic inflation, double-field chaotic inflation, and hybrid inflation.
For the first and second ones, we have (semi-)analytic methods for the squeezed bispectrum and thus compare them with the results of our method as a check, finding good agreement.
The third one is, as far as the authors know, the first numerical calculation of the bispectrum in the diffusion-dominated hybrid inflation with mild waterfall transition.
We actually obtain the bispectrum with significant magnitude, which may lead to nontrivial cosmological consequences.

In this paper, we adopt the Planck unit, $c=\hbar=\Mpl=1$ ($\Mpl$ is the reduced Planck mass).

\section{Preliminary \label{sec:prel}}

\subsection{Squeezed bispectrum of primordial perturbations}

For the gauge-invariant curvature perturbation $\zeta$, whose Fourier mode with the wavenumber $\bfk$ is given by
\bae{
\tilde{\zeta}(\bfk) \coloneqq \int \dd{\bfx} \zeta(\bfx) e^{-i \bfk \cdot \bfx},
}
the power spectrum $P_\zeta$ is defined via the two-point correlation function:
\bae{
\left<\tilde{\zeta}(\bfk_1)\tilde{\zeta}(\bfk_2)\right>=(2 \pi)^3 \delta^{(3)}(\bfk_1+\bfk_2) P_\zeta(k_1).
}
Here, $\left<~\right>$ denotes the ensemble average, and $P_\zeta$ depends only on $k_1$, the norm of the wavenumber vector $\bfk_1$, under the assumption of the statistical homogeneity and isotropy of the universe.
The dimensionless power spectrum is defined by
\bae{
\calP_\zeta(k) \coloneqq \frac{k^3}{2\pi^2} P_\zeta(k).
}

Similarly, the bispectrum $B_\zeta$ is defined via the three-point function:
\bae{
\left<\tilde{\zeta}(\bfk_1)\tilde{\zeta}(\bfk_2)\tilde{\zeta}(\bfk_3)\right>=(2 \pi)^3 \delta^{(3)}(\bfk_1+\bfk_2+\bfk_3) B_\zeta(k_1,k_2,k_3),
}
where the homogeneity and isotropy assumption makes $B_\zeta$ depend only on the norms $k_1$, $k_2$, and $k_3$.
Often, and also in this paper, we focus on $B_\zeta$ in the squeezed limit, where one of $k_1$, $k_2$, and $k_3$ is much smaller than the others, e.g., $k_1 \ll k_2 \simeq k_3$.
Hereafter, we denote the former by $k_\rmL$ and make an approximation that $k_2 = k_3 \coloneqq k_\rmS$, and regard $B_\zeta$ in this limit as a function only of $k_\rmL$ and $k_\rmS$: $B_\zeta=B_\zeta(k_\rmL,k_\rmS)$. 
We define its dimensionless version by
\begin{align}
    \calB_\zeta(k_\rmL,k_\rmS) \coloneqq \frac{k_\rmL^3}{2\pi^2}\frac{k_\rmS^3}{2\pi^2}B_\zeta(k_\rmL,k_\rmS).
\end{align}
The following non-linearity parameter $f_{\rm NL}$ is also often used to denote the magnitude of the bispectrum relative to the power spectrum:
\begin{align}
    \frac{3}{5}f_{\rm NL}(k_\rmL,k_\rmS) \coloneqq \frac{\mathcal{B}_\zeta(k_{\rm L},k_{\rm S})}{4\mathcal{P}_\zeta(k_{\rm L})\mathcal{P}_\zeta(k_{\rm S})}.
\end{align}

Ref.~\cite{Tada_2017} gives a formula for the bispectrum, on which the discussion in this paper relies.
We consider a patch of comoving size $k_\rmL^{-1}$ and define $\zeta_\rmL$, the coarse-grained curvature perturbation on this scale, by 
\bae{
\zeta_\rmL(\bfx) \coloneqq \int \dd{\bfy} W(k_\rmL|\bfx-\bfy|)\zeta(\bfy)
}
with the window function
\bae{
W(x) \coloneqq
\bce{
1 &; ~ x \le 1, \\
0 &; ~ \text{otherwise.}
}
}
We also consider the power spectrum of scale $k_\rmS$ evaluated within this patch, which we denote by $\calP_\zeta^{\rm patch}(k_\rmS)$.
Then, the bispectrum is given by the correlation of $\zeta_\rmL$ and $\calP_\zeta^{\rm patch}(k_\rmS)$ differentiated by $\log k_\rmL$:
\bae{\label{eq:calBTV}
\calB_\zeta(k_\rmL,k_\rmS) \simeq \dv{\log k_\rmL} \left<\zeta_\rmL \calP_\zeta^{\rm patch}(k_\rmS)\right>.
}

\subsection[$\delta N$ formalism]{\boldmath $\delN$ formalism}

Inflation occurs while the scalar fields called inflatons slowly roll in the flat part of their potential, and thus the potential is dominant in the energy density, ending when it becomes no longer dominant.
According to the $\delta N$ formalism~\cite{Starobinsky:1985ibc,Salopek:1990jq,Sasaki:1995aw,Wands:2000dp,Lyth:2004gb}, the curvature perturbation $\zeta$ is related to the fluctuation of the e-fold number $N\coloneqq\ln a$ taken during inflation, where $a$ is the scale factor.
More strictly, $\zeta$ is given by the fluctuation of the e-fold realised between the initial flat hypersurface and the final uniform density hypersurface, the \ac{EOI} surface:
\bae{
\zeta(\bfx) = \delcalN(\bfx) \coloneqq \calN(\bfx) - \bar{N},
}
where $\calN(\bfx)$ is the e-fold spent at the point $\bfx$ during inflation, and $\bar{N}$ is its ensemble average independent of $\bfx$ under the homogeneity assumption.

In the $\delN$ formalism, the curvature perturbation can be viewed as the perturbation of the e-fold number induced by the quantum fluctuation of the inflaton field values.
Concretely, this paper considers $d$ canonical scalar fields $\bfphi=(\phi_1, \ldots, \phi_d)$ minimally coupled to gravity and described by the action
\begin{align}
    S = \int\dd[4]{x}\sqrt{-g}\left[\frac{1}{2}R-\frac{1}{2}\sum_{i=1}^dg^{\mu\nu}\partial_\mu\phi_i\partial_\nu\phi_i-V(\bfphi)\right],
    \label{eq:action}
\end{align}
where $g_{\mu\nu}$ is the spacetime metric, $R$ is the Ricci scalar associated with it, and $V(\bfphi)$ is the potential of $\bfphi$.
Then, in cases where the usual slow-roll approximation is valid, $\zeta$ is given by a perturbative expansion of the e-fold number around the background field value trajectory, namely the slow-roll attractor $\bfphi_*$:
\bae{
\zeta \simeq \sum_{i=1}^d \frac{\partial N}{\partial \phi_i}(\bfphi_*) \delta \phi_i,
}
where $N(\bfphi)$ is the e-folds elapsed on the trajectory starting from the value $\bfphi$ and ending at the \ac{EOI} surface, and $\delta \phi_i$ is the quantum fluctuation of $\phi_i$.
To get $\calP_\zeta$, we can combine this with the power spectrum of the field fluctuations at Hubble exit time given by
\bae{
\calP_{\delta \phi}(k)=\eval{\left(\frac{H}{2\pi}\right)^2}_{k=aH}.
}
Here, the Hubble parameter $H\coloneqq \frac{\dot{a}}{a}$ is given via the Friedmann equation
\begin{equation}
    3H^2(\bfphi, \boldsymbol{\pi}) = \frac{1}{2} \boldsymbol{\pi}^2 + V(\bfphi),
\end{equation}
where $\boldsymbol{\pi}\coloneqq(\pi_1,\ldots,\pi_n)$ is $\bfphi$'s ($a^3$-rescaled) conjugate momenta in cosmic time $t$, and approximated by $3H^2 \simeq V$ in slow-roll.
We then reach the well-known formula
\bae{
\calP_\zeta(k) \simeq \calP_\zeta(\bfphi_*(k)) \qc \calP_\zeta(\bfphi) \coloneqq \frac{V(\bfphi)}{24 \pi^2 \epsilon_V(\bfphi)} \qc \epsilon_V(\bfphi) \coloneqq \frac{\sum_{i=1}^d \left(\partial_{\phi_i}V(\bfphi)\right)^2}{2\left(V(\bfphi)\right)^2},
}
where $\bfphi_*(k)$ is the field value on the background trajectory at the Hubble exit of scale $k$.

\begin{figure}
\centering
\includegraphics[width=0.7\linewidth]{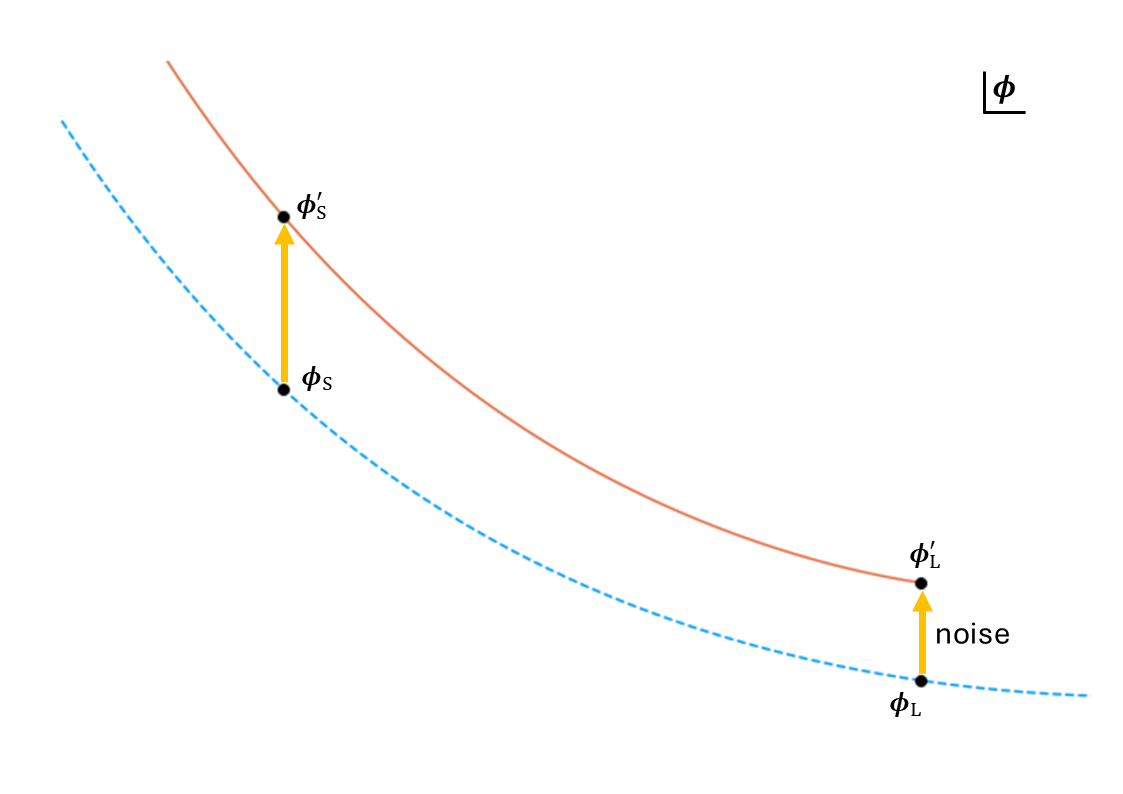}
\caption{Schematic illustration of how curvature perturbations of large and small scales correlate via the change of the path. The blue dashed curve is the slow-roll trajectory. Assuming that the field value on the path at the time of Hubble exit of the long scale is shifted, we show the trajectory starting from this shifted value in solid red.}
\label{fig:HybTraj}
\end{figure}

An intuition on how the bispectrum arises can also be obtained.
As implied by Eq.~\eqref{eq:calBTV}, the bispectrum arises via the correlation between long- and small-scale perturbations.
Such a correlation is induced through the change of the field value trajectory by the field fluctuation, which is schematically described in Figure~\ref{fig:HybTraj}.
Suppose that on the slow-roll trajectory, the field value fluctuates at the time of Hubble exit of the long scale $k_\rmL$.
This fluctuation yields the long-scale curvature perturbation $\zeta_\rmL$ and, additionally, changes the subsequent trajectory, including the field value at the time of Hubble exit of the small scale $k_\rmS$, which affects small-scale perturbations generated around that time.
This means that large-scale perturbations affect small-scale ones, which leads to their correlation.

Ref.~\cite{Tada_2017} provides a similar but more rigorous discussion to find the bispectrum.
They introduce two formalisms, the forward and backward formalisms, which can be switched by a local rescaling of the physical wavelengths.
In the forward formalism, the patchwise small-scale power spectrum $\calP_\zeta^{\rm patch}(k_\rmS)$ is evaluated at the time $\NL-\NS$ e-folds after the Hubble exit of the long scale $k_\rmL$, where $\NL$ and $\NS$ are defined via $k_\rmL=e^{-\NL}k_\mathrm{EOI}$ and $k_\rmS=e^{-\NS}k_\mathrm{EOI}$, respectively, with the scale $k_\mathrm{EOI}$ exiting the Hubble radius at the \ac{EOI}.
In the backward formalism, $\calP_\zeta^{\rm patch}(k_\rmS)$ is evaluated at the time $\NS$ e-folds before the \ac{EOI}.
Ref.~\cite{Tada_2017} then rederives the bispectrum in the single-field slow-roll case, which satisfies Maldacena’s \ac{CR}~\cite{Maldacena_2003}
\bae{
\frac{3}{5}f_{\rm NL}(k_\rmL,k_\rmS)=\frac{3}{2}\epsilon_V(\phi_*(k_\rmS))-\frac{1}{2}\eta_V(\phi_*(k_\rmS))=\frac{1-n_\rmS(k_\rmS)}{4}
}
in the forward formulation, where $\eta_V\coloneqq\frac{V^{\prime\prime}}{V}$ and $n_\rmS(k) \coloneqq \dv{\log \calP_\zeta(k)}{\log k}$, 
while it vanishes in the backward formulation, implying that the squeezed bispectrum is merely a consequence of the scale shift induced by the long-wavelength curvature perturbation $\zeta_\rmL$.

A similar discussion also applies to multi-field and/or diffusion-dominated cases.
In particular, in multi-field cases, a field fluctuation can cause a nonlinear shift of the subsequent trajectory, which may lead to the bispectrum being different from the \ac{CR}-based one.

\subsection{Stochastic formalism}

In the stochastic formalism of inflation, we probabilistically analyse the dynamics of the coarse-grained inflaton field values as described below.
We consider the system with the action~\eqref{eq:action}.
We use the flat slicing for the equal-time hypersurface and regard the e-folds $N$ as the time variable, with its initial value $N_0$.\footnote{See, e.g., Refs.~\cite{Finelli:2008zg,Finelli:2010sh,Pattison:2019hef} for discussions about the e-folding number as the time variable in the stochastic formalism.}
We denote the superHubble parts of $\bfphi$ and $\bfpi$ by $\boldsymbol{\varphi}=(\varphi_1,\ldots,\varphi_d)$ and $\boldsymbol{\varpi}=(\varpi_1,\ldots,\varpi_d)$, respectively, which are defined by coarse-graining:
\beae{
    &\varphi_i(N,\mathbf{x})\coloneqq\int\frac{\dd[3]{k}}{(2\pi)^3} e^{i \mathbf{k}\cdot\mathbf{x}} \tilde{\phi}_i(N,\mathbf{k})\Theta(k_\sigma(N)-k), \\
    &\varpi_i(N,\mathbf{x})\coloneqq\int\frac{\dd[3]{k}}{(2\pi)^3} e^{i \mathbf{k}\cdot\mathbf{x}} \tilde{\pi}_i(N,\mathbf{k})\Theta(k_\sigma(N)-k).
}
Here,
\bae{
\tilde{\phi}_i(N,\mathbf{k}) \coloneqq \int\dd[3]{x}e^{-i \mathbf{k}\cdot\mathbf{x}} \phi_i(N,\mathbf{x}) \qc \tilde{\pi}_i(N,\mathbf{k}) \coloneqq \int\dd[3]{x}e^{-i \mathbf{k}\cdot\mathbf{x}} \pi_i(N,\mathbf{x})
}
are the Fourier modes of $\phi_i$ and $\pi_i$, respectively.
\bae{
\Theta(z) =
\bce{
1 & ; ~ z>0, \\
0 & ; ~ z<0,
}
}
is the Heaviside function.\footnote{The consistent definition of $\Theta(z)$ at $z=0$ requires a detailed discussion of the discretisation of the path
integral. See Refs.~\cite{Tokuda:2017fdh,Tokuda:2018eqs}.}
$k_\sigma(N)$ is defined by $k_\sigma(N)\coloneqq \sigma a(N) \mathsf{H}$ with the dimensionless and dimensionful parameters $\sigma$ and $\mathsf{H}$, which are set so that $k_\sigma(N) \ll a(N) H(\boldsymbol{\varphi}(N), \boldsymbol{\varpi}(N))$ holds for typical sample paths in the range of $N$ under consideration.\footnote{The coarse-graining scale $k_\sigma$ is often defined by $\sigma a(N)H(\bm{\varphi}(N),\bm{\varpi}(N))$ in the literature, but it requires a circular definition of $\bm{\varphi}$ and $\bm{\varpi}$. In this work, we rather use a certain constant $\sfH$ as a model parameter. It is less physical but can avoid the circular definition, and the original algorithm for the power spectrum proposed in Ref.~\cite{Ando:2020fjm} needs to suppose that the Hubble parameter is almost constant. See also footnote~3 of Ref.~\cite{Mizuguchi:2024kbl}.}
In this paper, we set $\sigma=0.1$ and $\mathsf{H}=H(\bfphi_{\rm ini}, \boldsymbol{\pi}_{\rm ini})$, assuming that $(\bfphi,\boldsymbol{\pi})$ takes a globally equal initial value $(\bfphi_{\rm ini},\boldsymbol{\pi}_{\rm ini})$.
With these setups, the time evolution of $\boldsymbol{\mathcal{X}}\coloneqq(\boldsymbol{\varphi},\boldsymbol{\varpi})$ is described by the Langevin equation
\beae{
    &\dv{\varphi_i(N)}{N}=\frac{\varpi_i (N)}{H(\boldsymbol{\varphi}(N), \boldsymbol{\varpi}(N))} + \xi_{\phi_i}(N), \\
    &\dv{\varpi_i(N)}{N}=-3\varpi_i(N)-\frac{\partial_{\phi_i}V(\boldsymbol{\varphi}(N))}{H(\boldsymbol{\varphi}(N), \boldsymbol{\varpi}(N))}+\xi_{\pi_i}(N),
    \label{eq:SDE}
}
with the initial value $\bfX_\ini=(\bfphi_{\rm ini},\boldsymbol{\pi}_{\rm ini})$.
Here, $\xi_X$ ($X=\phi_i$ or $\pi_i$) are the white Gaussian noises with zero means and covariances
\begin{align}\label{eq: noise correlation}
    \expval{\xi_X(N)\xi_Y(N')}=\calP_{XY}(k_\sigma(N))\delta(N-N'),
\end{align}
for $X,Y=\phi_i,\pi_i$, where the power spectrum $\mathcal{P}_{XY}$ defined by
\bae{\label{eq: calP}
    \expval{\hat{X}_\bfk\hat{Y}_{\bfk'}}=(2\pi)^3\delta^{(3)}(\bfk+\bfk')\frac{2\pi^2}{k^3}\calP_{XY}(k),
}
with the corresponding quantum operators $\hat{X}_\bfk$ and $\hat{Y}_\bfk$ in Fourier space.\footnote{$\expval{\cdot}$ represents the quantum average in Eq.~\eqref{eq: calP}, although it denotes the ensemble average in other places.}
Note that although $\boldsymbol{\mathcal{X}}$ (and also $\boldsymbol{\xi}$) takes different values at different spatial points $\mathbf{x}$, we have omitted $\mathbf{x}$ in Eq.~\eqref{eq:SDE}.
This is because we will hereafter solve Eq.~\eqref{eq:SDE} without considering $\mathbf{x}$ dependence, and regard each sample path as representing a realisation of the time-evolving $\boldsymbol{\mathcal{X}}(N)$ at one spatial point.

We need to discretise Eq.~\eqref{eq:SDE} in time to generate a sample path of $\boldsymbol{\mathcal{X}}$.
In this work, we adopt the \ac{EM} method~\cite{Maruyama1955}: setting grid points $N_0,N_1,N_2,\ldots$ in time with interval $\Delta N=N_{m+1}-N_m$, we iteratively calculate $\bfcalX(N_1),\bfcalX(N_2),\ldots$ by
\beae{
    &\bmte{\varphi_i(N_{m+1}) =\varphi_i(N_m) + \frac{\varpi_i(N_m)}{H(\boldsymbol{\varphi}(N_m), \boldsymbol{\varpi}(N_m))}\Delta N
    +\mathcal{P}_\phi(N_m,\boldsymbol{\varphi}(N_m), \boldsymbol{\varpi}(N_{m+1})) \Delta W_{i,m}, } \\
    &\bmte{\varpi_i(N_{m+1}) = \varpi_i(N_m) + \left(-3\varpi_i(N_m)-\frac{\partial_{\phi_i} V(\boldsymbol{\varphi}(N_m))}{H(\boldsymbol{\varphi}(N_m), \boldsymbol{\varpi}(N_m))}\right) \Delta N,}
    \label{eq:SDEDisc}
}
where $\Delta W_{i,m}$'s are independent samples from the normal distribution with zero mean and variance $\Delta N$.
Here and hereafter, we neglected $\xi_{\pi_i}$ as they are slow-roll suppressed in general~\cite{Fujita_2025}.

In the stochastic-$\delN$ formalism, the combination of this stochastic formalism and the $\delN$ formalism, a formula for the power spectrum is presented by Ref.~\cite{Ando:2020fjm}.
We define a random variable $\mathcal{N}_\mathbf{X}$ as the first time when a path of $\boldsymbol{\mathcal{X}}$ starting from $\mathbf{X}$ reaches the \ac{EOI} surface.
We also define
\begin{align}
\delta \mathcal{N}_{\mathbf{X}}\coloneqq\mathcal{N}_{\mathbf{X}}-\left\langle 
\mathcal{N}_{\mathbf{X}} \right\rangle.
\end{align}
Then, we can calculate $\mathcal{P}_\zeta$ by the formula in Ref.~\cite{Ando:2020fjm}, which is rewritten by Ref.~\cite{miyamoto2025calculating} as follows:
\beae{\label{eq:PSAndo}
    &\mathcal{P}_\zeta(k) \simeq \dv{F_{\langle\delta \mathcal{N}^2\rangle}}{N_\mathrm{bk}}{}(N_\mathrm{bk}(k)), \\
    &F_{\langle\delta \mathcal{N}^2\rangle}(N_\mathrm{bk}) \coloneqq \int \dd{\mathbf{X}} P_\mathrm{bk}\left(N_\mathrm{bk},\mathbf{X};\bfX_\ini\right) \left\langle \delta \mathcal{N}_{\mathbf{X}}^2 \right\rangle.
}
Here, $P_\mathrm{bk}(N_\mathrm{bk},~\cdot~;\bfX_0)$ is the \ac{PDF} of $\bfcalX$ starting from a value $\bfX_0$ at backward e-fold $N_\mathrm{bk}$, that is, the \ac{PDF} at the time $N_\mathrm{bk}$ e-folds before the \ac{EOI}.
\bae{\label{eq:Nbk}
N_{\mathrm{bk}}(k)\coloneqq-\log \left(\frac{k}{k_{\mathrm{end}}}\right)
}
is the backward e-fold at which the scale $k$ exits the coarse-graining radius, and $k_{\rm end}\coloneqq k_\sigma(N_{\rm end})$ denotes the coarse-graining scale at the \ac{EOI}. 
We can regard $\mathcal{P}_\zeta$ as a function of $N_\mathrm{bk}$, which is common in the context of the stochastic-$\delN$ formalism~\cite{Fujita:2014tja,Kawasaki:2015ppx,Tada:2023fvd}, letting $\calP_\zeta(N_\mathrm{bk})$ be the value of Eq.~\eqref{eq:PSAndo} with $N_{\mathrm{bk}}(k)$ replaced by $N_\mathrm{bk}$.

Ref.~\cite{miyamoto2025calculating} proposed a Monte Carlo-based algorithm for calculating Eq.~\eqref{eq:PSAndo}.
Leaving the details to Ref.~\cite{miyamoto2025calculating}, we present the points of the method.
We observe that $\expval{\delta\calN_\bfX^2}$, the variance of $\calN_\bfX$, is given by
\bae{\label{eq:delNxEst}
\expval{\delta\calN_\bfX^2} = \expval{\frac{1}{2}\left(\calN_\bfX^{(1)}-\calN_\bfX^{(2)}\right)^2},
}
where $\calN_\bfX^{(1)}$ and $\calN_\bfX^{(2)}$ are two independent samples of $\calN_\bfX$.
Thus, the following procedure gives $F_{\langle\delta \mathcal{N}^2\rangle}(N_\mathrm{bk})$, and numerically differentiating it gives $\calP_\zeta(N_\bk)$~:
\begin{enumerate}
    \item Generate a path of $\bfcalX$, and on this path, take the value $\bfX_\bk$ at which the backward e-fold is $N_\bk$.
    \item Generate two paths from $\bfX_\bk$, and let the total e-folds of these paths be $N_1$ an $N_2$.
    \item Repeat 1 and 2, and calculate the average of $\frac{1}{2}(N_1-N_2)^2$.
\end{enumerate}

\section{Proposed method \label{sec:method}}

Now, we present a Monte Carlo-based algorithm to calculate $\mathcal{B}_\zeta$.
Hereafter, as is $\calP_\zeta$, we regard $\mathcal{B}_\zeta$ as a function of the backward e-folds $\NL$ and $\NS$, letting $\mathcal{B}_\zeta(\NL,\NS)$ be the value of $\mathcal{B}_\zeta(k_\rmL,k_\rmS)$ with $k_\rmL$ and $k_\rmS$ such that $N_{\mathrm{bk}}(k_\rmL)=\NL$ and $N_{\mathrm{bk}}(k_\rmS)=\NS$, respectively.

We begin with considering the meaning of ``evaluating $\zeta_\rmL$ and $\calP_\zeta^{\rm patch}(\NS)$ within the path" and ``taking the expectation of $\zeta_\rmL \calP_\zeta^{\rm patch}(\NS)$" in Eq.~\eqref{eq:calBTV}.
Consider the last time at which all the comoving points in the patch lie within the coarse-graining radius $k_\sigma^{-1}$, which we hereafter call {\it decoupling}.
$\bfcalX$ at that time, which is denoted by $\bfcalX_{\rm dec}$, is the same at every point in patch.
Thus, we consider that the power spectrum evaluated within the patch is calculated via the expectation conditional on $\bfcalX_{\rm dec}$:
\beae{\label{eq:calPPatchApp}
&\calP_\zeta^{\rm patch}(\NS) \simeq \calP_\zeta^{\rm patch}(\NS\mid\bfcalX_{\rm dec}) = \dv{\NS}F_{\langle\delta \mathcal{N}^2\rangle}(\NS \mid \bfcalX_{\rm dec}), \\
&F_{\langle\delta \mathcal{N}^2\rangle}(\NS \mid \bfcalX_{\rm dec}) \coloneqq \int \dd \bfXS P_\rmS\left(\NS,\bfXS\mid\bfcalX_{\rm dec}\right) \left\langle \delta \mathcal{N}_{\bfXS}^2 \right\rangle.
}
Here, the \ac{PDF} $P_\rmS\left(\NS,~\cdot~\mid\bfcalX_{\rm dec}\right)$ conditional on $\bfcalX_{\rm dec}$ is chosen according to whether we take the forward or backward formulation. In the forward formulation, it is the conditional PDF of $\bfcalX$ $\NL-\NS$ e-fold after the decoupling given $\bfcalX_{\rm dec}$, which is, because of the Markovianity of $\bfcalX$,
\bae{
P_\rmS\left(\NS,\bfXS\mid\bfcalX_{\rm dec}\right) = P_\mathrm{fw}\left(\NL-\NS,\bfXS;\bfcalX_{\rm dec}\right),
}
where $P_\mathrm{fw}\left(N,~\cdot~;\bfX_0\right)$ is the \ac{PDF} of $\bfcalX$ $N$ e-folds after its start from $\bfX_0$.
In the backward formulation, $P_\rmS\left(\NS,~\cdot~\mid\bfcalX_{\rm dec}\right)$ is the conditional PDF of $\bfcalX$ at backward e-fold $\NS$ given $\bfcalX_{\rm dec}$, which is, because of the Markovianity again,
\bae{
P_\rmS\left(\NS,\bfXS\mid\bfcalX_{\rm dec}\right) = P_\mathrm{bk}\left(\NS,\bfXS;\bfcalX_{\rm dec}\right),
}
the \ac{PDF} of $\bfcalX$ starting from $\bfcalX_{\rm dec}$ at backward e-fold $\NS$.
For $\zeta_\rmL$, assuming that the spatial average in its definition is approximated by the ensemble average, we regard $\zeta_\rmL$ as
\bae{\label{eq:zetaLApp}
\zeta_\rmL \simeq \zeta_\rmL(\bfcalX_{\rm dec}) = \int \dd N P_\calN(N\mid\bfcalX_{\rm dec}) (N-\bar{N}),
}
where $P_\calN(~\cdot~\mid\bfcalX_{\rm dec})$ is the PDF of the e-fold elapsed during inflation at points in the patch conditional on $\bfcalX_{\rm dec}$.
Then, we consider that
\bae{\label{eq:zetaLcalPExp}
\expval{\zeta_\rmL \calP_\zeta^{\rm patch}(\NS)} = \int \dd{\bfX_{\rm dec}} P_{\bfcalX_{\rm dec}}(\bfX_{\rm dec}) \zeta_\rmL(\bfX_{\rm dec}) \calP_\zeta^{\rm patch}(\NS\mid\bfX_{\rm dec}),
}
namely, the expectation with respect to the randomness of $\bfcalX_{\rm dec}$, where $P_{\bfcalX_{\rm dec}}$ is the \ac{PDF} of $\bfcalX_{\rm dec}$.
Plugging Eq.~\eqref{eq:zetaLApp} into Eq.~\eqref{eq:zetaLcalPExp} yields
\bae{\label{eq:zetaLcalPExpAppTemp}
\expval{\zeta_\rmL \calP_\zeta^{\rm patch}(\NS)} & \simeq \int \dd{N} \dd{\bfX_{\rm dec}} P_{\bfcalX_{\rm dec}}(\bfX_{\rm dec}) P_\calN(N\mid\bfX_{\rm dec}) (N-\bar{N}) \calP_\zeta^{\rm patch}(\NS\mid\bfX_{\rm dec}) \nonumber \\
& = \int \dd{N} \dd{\bfX_{\rm dec}} P_{\calN,\bfcalX_{\rm dec}}(N,\bfX_{\rm dec})  (N-\bar{N}) \calP_\zeta^{\rm patch}(\NS\mid\bfX_{\rm dec}),
}
where $P_{\calN,\bfcalX_{\rm dec}}$ is the joint \ac{PDF} of $\calN$ and $\bfcalX_{\rm dec}$.
We further make an approximation with respect to $\bfcalX_{\rm dec}$: we identify it with $\bfcalX$ at backward e-fold $\NL$, which we denote by $\bfcalX_\rmL$. 
Note that at each comoving point in the patch, the time $\NL$ e-folds before the \ac{EOI} is not the time of decoupling, because the total e-fold elapsed during inflation differs at the points.
Nevertheless, as long as the magnitude of the curvature perturbation is small, we can approximately regard these two times as equal, and the above identification is expected to be valid.
We then replace $P_{\calN,\bfcalX_{\rm dec}}$ in Eq.~\eqref{eq:zetaLcalPExpAppTemp} with $P_{\calN,\bfcalX_\rmL}$, the joint \ac{PDF} of $\calN$ and $\bfcalX_\rmL$ as\footnote{A similar approximation is also used in Ref.~\cite{Ando:2020fjm}; see Sec.~3.3 in it.}
\bae{\label{eq:zetaLcalPExpApp}
\expval{\zeta_\rmL \calP_\zeta^{\rm patch}(\NS)} \simeq \int \dd{N} \dd{\bfX_\rmL} P_{\calN,\bfcalX_\rmL}(N,\bfX_\rmL)  (N-\bar{N}) \calP_\zeta^{\rm patch}(\NS\mid\bfX_\rmL).
}
By combining Eqs.~\eqref{eq:calBTV}, \eqref{eq:delNxEst}, \eqref{eq:calPPatchApp}, and \eqref{eq:zetaLcalPExpApp}, and recalling the relationship in Eq.~\eqref{eq:Nbk}, we finally get
\beae{\label{eq:calBFormula}
\calB_\zeta(\NL,\NS) & \simeq - \frac{\partial^2 F_{\calB_\zeta}}{\partial \NL \partial \NS} (\NL,\NS), \\
F_{\calB_\zeta}(\NL,\NS) & \bmte{\coloneqq \int \dd{N} \dd{\bfX_\rmL} \dd{\bfX_\rmS} \dd{N_1} \dd{N_2} P_{\calN,\bfcalX_\rmL}(N,\bfX_\rmL) P_\rmS\left(\NS,\bfX_\rmS\mid\bfX_\rmL\right) \\
\times P_{\calN_{\bfX_\rmS}}(N_1) P_{\calN_{\bfX_\rmS}}(N_2) (N-\bar{N}) \frac{1}{2}\left(N_1-N_2\right)^2,}
}
where $P_{\calN_{\bfX_\rmS}}$ is the \ac{PDF} of $\calN_{\bfX_\rmS}$, the e-fold elapsed on a path of $\bfcalX$ starting from the value $\bfX_0$ and ending on the \ac{EOI} surface.

We can estimate $F_{\calB_\zeta}(\NL,\NS)$ by the Monte Carlo method, that is, sampling the random variables that appear in Eq.~\eqref{eq:calBFormula} and taking the average of the sample values of the integrand.
In fact, the sampling can be done as follows.
We can get a sample value $(N,\bfX_\rmL)$ of $(\calN,\bfcalX_\rmL)$ by generating a path from $\bfX_\ini$ to the \ac{EOI} surface and finding the total e-fold of the path and the value of $\bfcalX$ on the path at backward e-folds $\NL$.
Given $\bfX_\rmL$, the sample value $\bfX_\rmS$ is obtained as follows: we generate a path starting from $\bfX_\rmL$ and ending at the \ac{EOI} surface, and then, as $\bfX_\rmS$, we take the value of $\bfcalX$ $\NL-\NS$ e-folds after the start (resp. $\NS$ e-folds before the end) on the path in the forward (resp. backward) formalism.
The two samples $N_1$ and $N_2$ of $\calN_{\bfX_\rmS}$ are obtained by generating two independent paths from $\bfX_\rmS$ to the \ac{EOI} surface and finding their total e-folds.
$\bar{N}$ can be obtained in advance by generating many paths from $\bfX_\ini$ to the \ac{EOI} surface and averaging their total e-folds.

\begin{algorithm} 
    \caption{The proposed method for calculating the squeezed bispectrum by Monte Carlo simulation}
    \label{alg:main}
    \begin{algorithmic}[1]

    \Require
    \Statex
    \begin{itemize}
        \item $N_{\rm samp}$: the number of samples
        \item $\mathbf{X}_\mathrm{ini}$: the initial values of $\boldsymbol{\mathcal{X}}$
        \item $N_{\mathrm{L}}, N_{\mathrm{S}}$: the backward e-folds that correspond to the scales at which the squeezed bispectrum is calculated
        \item $\Delta N_{\rm bk}$: step size of the finite difference approximations of derivatives with respect to $N_{\mathrm{L}}$ and $N_{\mathrm{S}}$
    \end{itemize}

    \Ensure An estimate $\hat{\mathcal{B}}(N_{\mathrm{L}}, N_{\mathrm{S}})$ of $\mathcal{B}(N_{\mathrm{L}}, N_{\mathrm{S}})$

    \State Generate $N_{\rm samp}$ paths from $\mathbf{X}_\mathrm{ini}$ to the \ac{EOI} surface, and let the average of their total e-folds be $\bar{N}$.

    \For{$n=1,...,N_{\rm samp}$}

    \State Generate a path $\omega_n$ from $\mathbf{X}_\mathrm{ini}$ to the \ac{EOI} surface, and let its total e-fold be $N_n$.
    
    \State Let the values of $\boldsymbol{\mathcal{X}}$ at the backward e-folds $N_{\rm L} \pm \Delta N_{\rm bk}$ on $\omega_n$ be $\bfX_{\rmL\pm,n}$, respectively.

    \State Generate two paths $\omega_{\rmL\pm,n}$ from $\bfX_{\rmL\pm,n}$ to the \ac{EOI} surface with the same \ac{PRNG} seed.

    \If{the forward formulation is adopted}
    \State Let the values of $\boldsymbol{\mathcal{X}}$ $N$ e-folds after the start on $\omega_{\rmL\pm,n}$ be $\bfX_{\rmL\pm,\rmS\pm,n}$, where $N=(N_\rmL \pm \Delta N_{\rm bk})-(N_\rmS \pm \Delta N_{\rm bk})$.
    \ElsIf{the backward formulation is adopted}
    \State Let the values of $\boldsymbol{\mathcal{X}}$ at the backward e-folds $N_\rmS \pm \Delta N_{\rm bk}$ on $\omega_{\rmL\pm,n}$ be $\bfX_{\rmL\pm,\rmS\pm,n}$
    \EndIf

    \For{$l=1,2$}

    \State Generate four paths from $\bfX_{\rmL\pm,\rmS\pm,n}$ to the \ac{EOI} surface with the same \ac{PRNG} seed, and let their total e-folds be $N_{\rmL\pm,\rmS\pm,l,n}$, respectively.

    \EndFor
    \EndFor

    \State
    Output
    \bme{
    \hat{\mathcal{B}}(N_L,N_S) \coloneqq 
    \frac{-1}{N_{\rm samp} \cdot 8\Delta N_{\rm bk}^2} \\
    \times\sum_{n=1}^{N_{\rm samp}} \left(N_n-\bar{N}\right) \left[\left(\left(N_{\rmL+,\rmS+,1,n}-N_{\rmL+,\rmS+,2,n}\right)^2-\left(N_{\rmL+,\rmS-,1,n}-N_{\rmL+,\rmS-,2,n}\right)^2\right)\right. \\
    \left.-\left(\left(N_{\rmL-,\rmS+,1,n}-N_{\rmL-,\rmS+,2,n}\right)^2-\left(N_{\rmL-,\rmS-,1,n}-N_{\rmL-,\rmS-,2,n}\right)^2\right)\right]
    \label{eq:BisAppComSeed}
    }

    \end{algorithmic}
\end{algorithm}

Now, there is an issue: to get $\calB_\zeta$, we need to differentiate $F_{\calB_\zeta}$ by $\NL$ and $\NS$.
A simple idea is to use some numerical differentiation method such as the central difference method: we compute $F_{\calB_\zeta}$ for four input values $(N_L \pm \Delta N_\bk, N_S \pm \Delta N_\bk)$ with a small shift $\Delta N_\bk$ and approximate
\bme{\label{eq:FinDiff}
\frac{\partial^2 F_{\calB_\zeta}}{\partial \NL \partial \NS} (\NL,\NS) \\
\simeq \frac{1}{4\Delta N^2}\left(F_{\calB_\zeta}(\NL + \Delta N_\bk,\NS + \Delta N_\bk)-F_{\calB_\zeta}(\NL + \Delta N_\bk,\NS - \Delta N_\bk) \right. \\
\left.-F_{\calB_\zeta}(\NL - \Delta N_\bk,\NS + \Delta N_\bk)+F_{\calB_\zeta}(\NL - \Delta N_\bk,\NS - \Delta N_\bk)\right).
}
However, as is widely known, finite difference methods may yield large errors when the differentiated function is evaluated by the Monte Carlo method \cite{glasserman2004monte}, since the estimates of $F_{\calB_\zeta}(\NL \pm \Delta N_\bk,\NS \pm \Delta N_\bk)$ have statistical errors, and they are amplified when divided by small $\Delta N_\bk^2$.

To mitigate this issue, we use the technique called the \ac{CRN} method \cite{glasserman2004monte}.
In calculating $F_{\calB_\zeta}$, we generate many paths of $\bfcalX$ based on Eq.~\eqref{eq:SDEDisc}, where a set of random numbers $\{\Delta W_{i,m}\}$ is used.
In the \ac{CRN} method, we use the same set to calculate $F_{\calB_\zeta}$ for four input values.
This makes the statistical errors in the four values of $F_{\calB_\zeta}$ similar, resulting in their cancellation in Eq.~\eqref{eq:FinDiff} and the smaller error in $\frac{\partial^2 F_{\calB_\zeta}}{\partial \NL \partial \NS}$ than when using different sets of $\Delta W_{i,m}$.
In practice, to generate $\Delta W_{i,m}$, we use some \ac{PRNG}, such as Xoshiro256++ \cite{Blackman2021} used in the Julia package \texttt{DifferentialEquations.jl}~\cite{rackauckas2017differentialequations}, which we use in the numerical demonstrations later, and using common random numbers is realised by setting a common seed of the \ac{PRNG}.

\begin{figure}
	\centering
        \includegraphics[width=\linewidth]{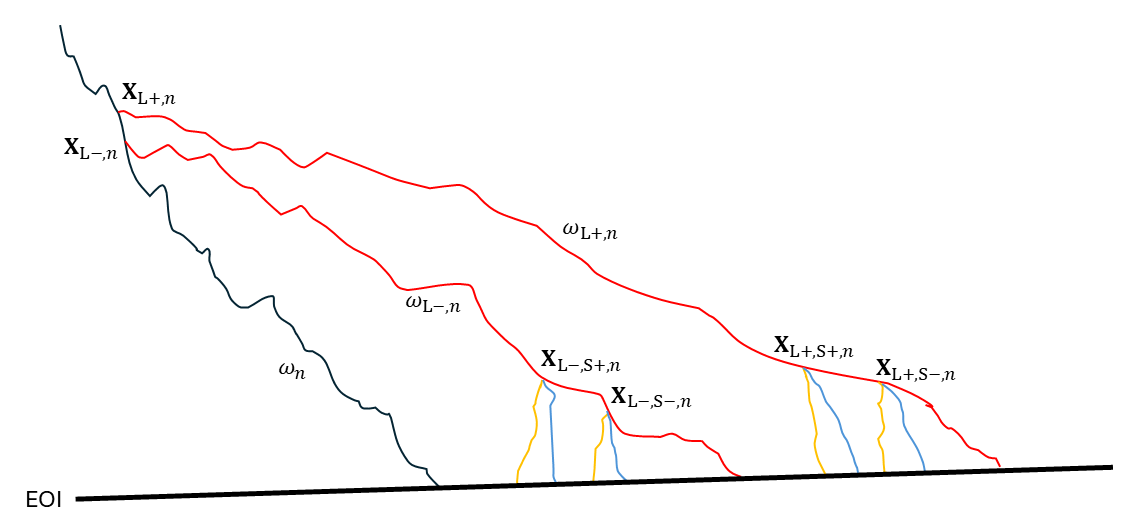}
	\caption{Schematic illustration of paths generated in Algorithm~\ref{alg:main}. The paths with the same colour are generated with the same \ac{PRNG} seed.}
    \label{fig:path}
\end{figure}

In summary, the proposed method for calculating $\calB_\zeta$ is as presented in Algorithm \ref{alg:main}.
Paths generated in this algorithm are schematically shown in Figure~\ref{fig:path}.

\section{Numerical demonstrations \label{sec:result}}

We now demonstrate the proposed method for some inflation models.

We use \texttt{DifferentialEquations.jl}~\cite{rackauckas2017differentialequations} to generate paths according to Eq.~\eqref{eq:SDEDisc}, adopting the \ac{EM} scheme with time step size $\Delta N=0.01$ and setting the number of samples to $N_{\rm samp}=10^6$.
All the calculations below were run on Fujitsu LIFEBOOK WP1/J3 with Intel Core Ultra 7 155H CPU (16 cores, 3.0 GHz), 32 GB RAM, and no GPU use.

\subsection{Single-field chaotic inflation}\label{sec: single chaotic}

\begin{figure}
\centering

\begin{subfigure}{0.49\linewidth}
\centering
\includegraphics[width=\linewidth]{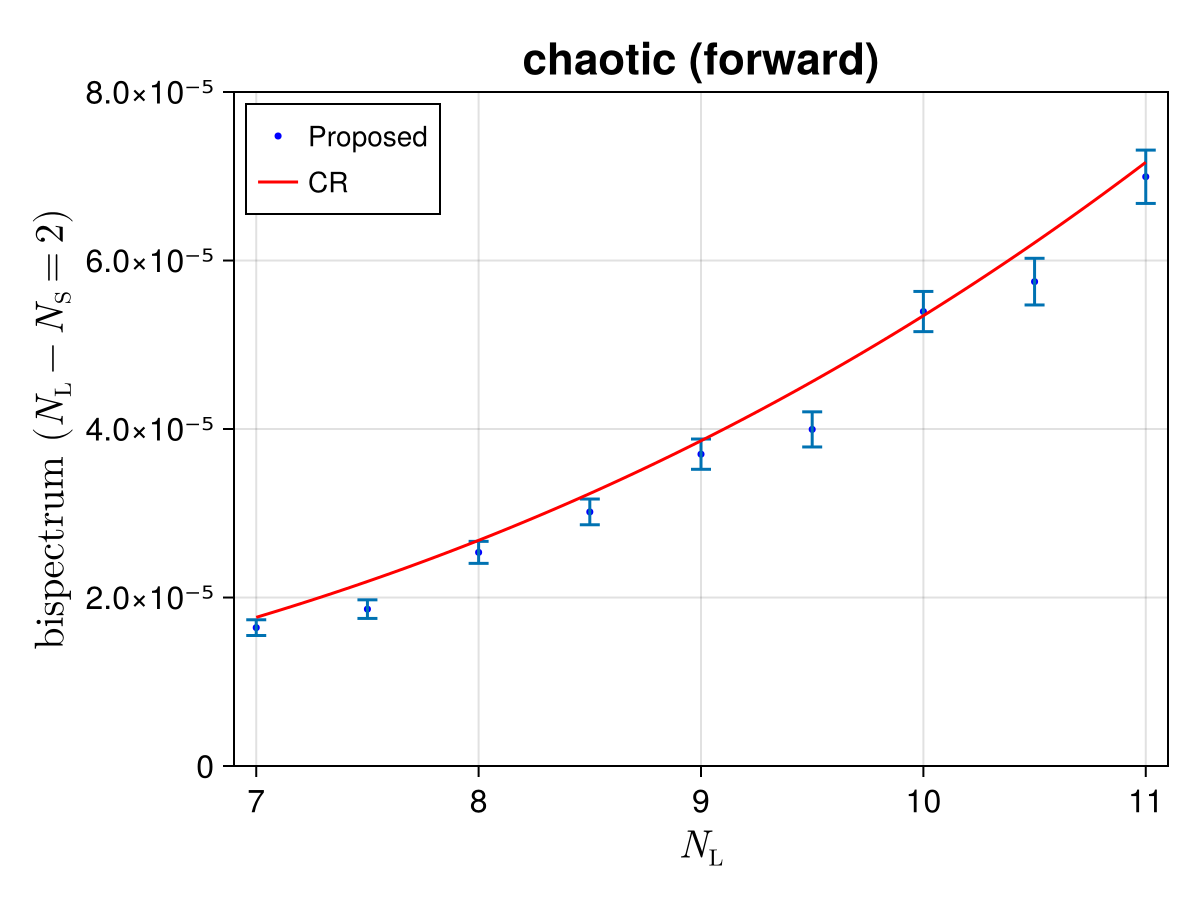}
\caption{Forward formulation}
\end{subfigure}
\hfill
\begin{subfigure}{0.49\linewidth}
\centering
\includegraphics[width=\linewidth]{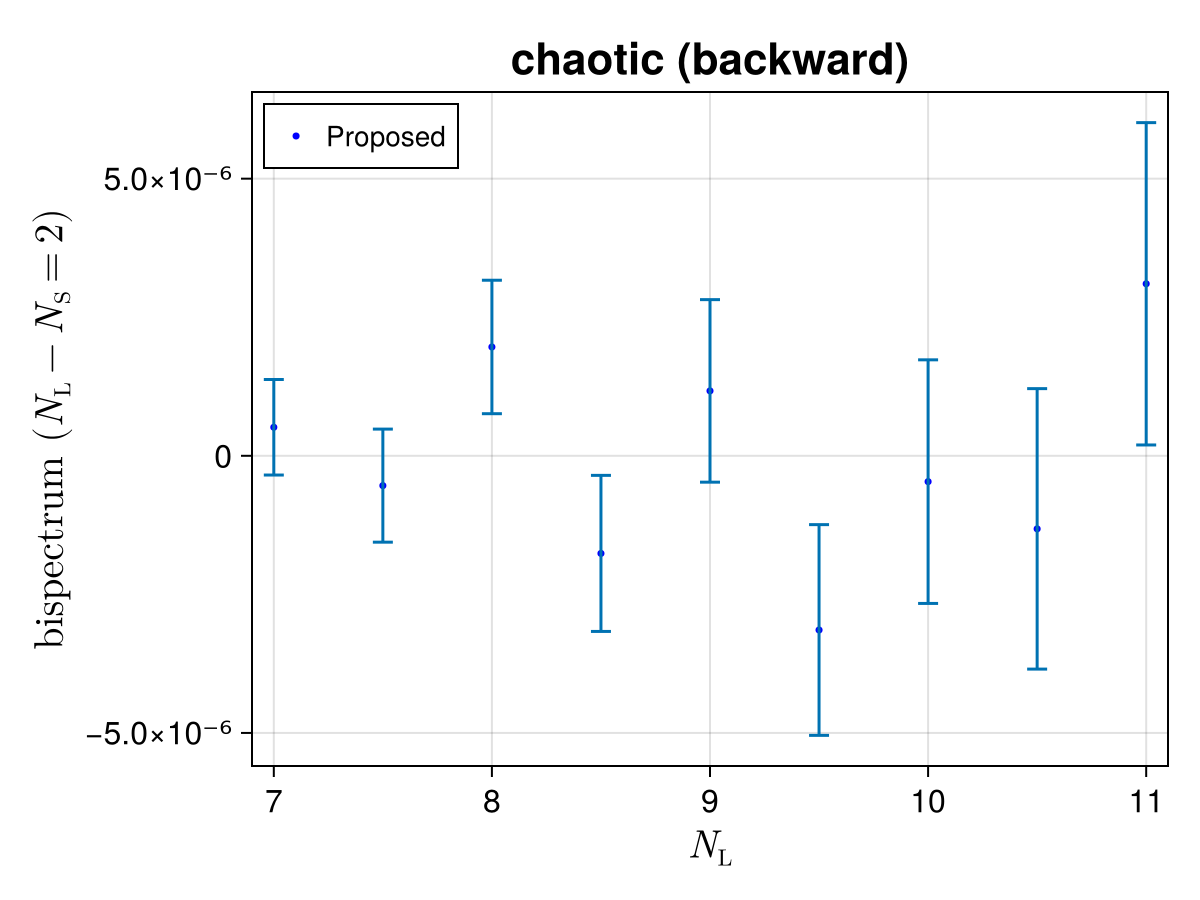}
\caption{Backward formulation}
\end{subfigure}

\caption{Bispectrum in the single-field chaotic inflation in the forward and backward formulation calculated by Algorithm \ref{alg:main} (blue points with error bars). In the forward formulation case, we also show the \ac{CR}-based approximation as the red line.}
\label{fig:chaotic}
\end{figure}

First, for a check of the proposed method, let us consider the model in which the analytic formula of the bispectrum is available, and we can compare the result of our method with it.
That is, we consider the single-field chaotic inflation~\cite{Linde:1983gd}, where the potential is given by
\bae{
V(\phi)=\frac{1}{2}m^2\phi^2,
}
with the inflaton mass $m$.
We set the model parameters as $m=0.1$, $\phi_{\rm ini}=8$, and $\pi_{\rm ini}=-\sqrt{\frac{2}{3}}m$, and define the \ac{EOI} surface by $\phi=\phi_{\rm end}\coloneqq\sqrt{2}$, at which the slow-roll parameters $\epsilon_V$ and $\eta_V$ becomes unity.

In Figure~\ref{fig:chaotic}, we plot $\calB(\NL,\NS)$ with $\NL-\NS=2$ calculated by Algorithm \ref{alg:main} along with the $1\sigma$ standard error in the Monte Carlo estimation shown as the error bar (the same applies to the graphs below).
In the forward formulation, we also show $\calB_\zeta$ analytically calculated via the \ac{CR}.
We find $\phi$ as a function of $N_\bk$ by
\bae{
\phi(N_\bk) \simeq \sqrt{\phi_{\rm end}^2 + 4N_\bk},
}
which follows from $\phi$'s equation of motion $3H^2 \dv{\phi}{N} \simeq -V^\prime$ and the Friedmann equation $3H^2 \simeq V$ under the slow-roll approximation, and $\epsilon_V(N_\bk)=\eta_V(N_\bk)=\frac{2}{\phi^2(N_\bk)}$.
We then calculate $\calB_\zeta(\NL,\NS)=\frac{12}{5} f_{\rm NL}(\NS)\calP_\zeta(\NL)\calP_\zeta(\NL)$ with $f_{\rm NL}(\NS) = \frac{3}{2}\epsilon_V(\NS)-\frac{1}{2}\eta_V(\NS)$ and $\calP_\zeta(N_\bk) \simeq \frac{V(\phi(N_\bk))}{24\pi^2\epsilon_V(N_\bk)}$.
On the other hand, the bispectrum in the backward formulation should be zero, as is the case for general single-field, slow-roll models~\cite{Tada_2017}.

Figure~\ref{fig:chaotic} implies that our method 
works well.
In the forward formulation, our method outputs the result that fits the analytical calculation.
In the backward formulation, our method's result is consistent with zero.

\subsection{Double-field chaotic inflation}\label{sec: double chaotic}

\begin{figure}[t]
\centering

\begin{subfigure}{0.49\linewidth}
\centering
\includegraphics[width=\linewidth]{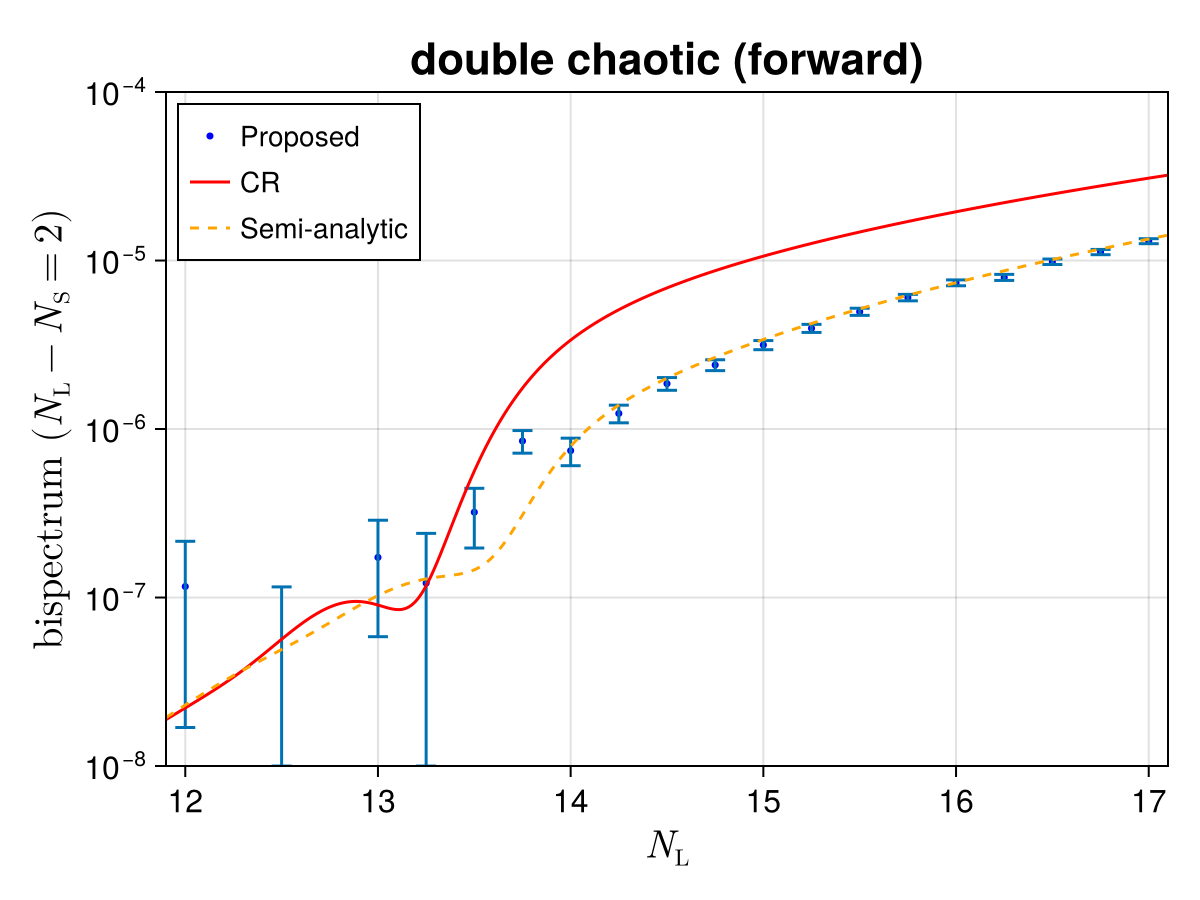}
\caption{Forward formulation}
\end{subfigure}
\hfill
\begin{subfigure}{0.49\linewidth}
\centering
\includegraphics[width=\linewidth]{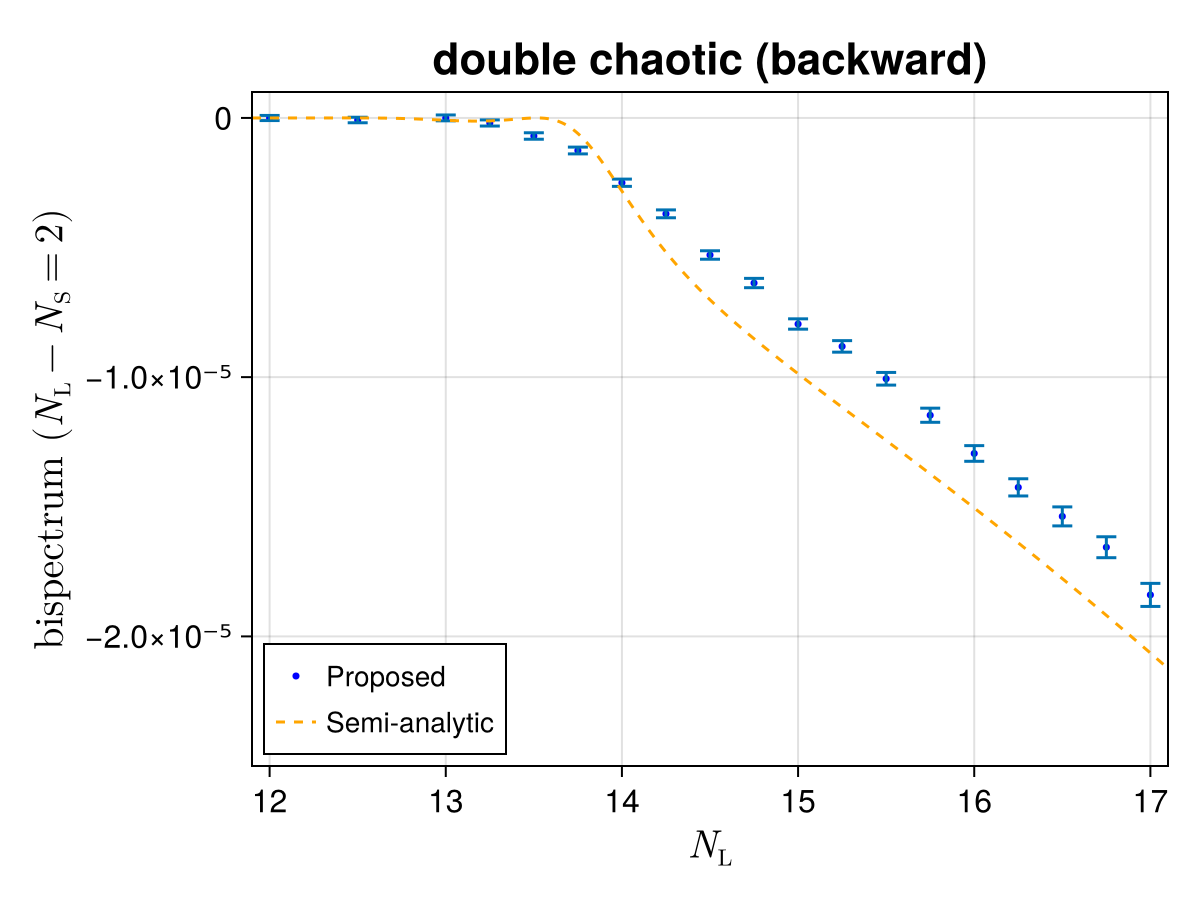}
\caption{Backward formulation}
\end{subfigure}

\caption{Bispectrum in the double-field chaotic inflation with masses $(m_\phi,m_\psi)=\left(0.1, \frac{0.1}{9}\right)$ in the forward and backward formulation calculated by Algorithm~\ref{alg:main} (blue points with error bars). We also show the result of the semi-analytic method in Ref.~\cite{Tada_2017} in dashed orange. In the forward formulation case, we also show the \ac{CR}-based bispectrum in solid red.}
\label{fig:doublechaotic}
\end{figure}

The second example is also for the check.
We consider inflation with two fields $(\phi,\psi)$, whose potential has a separable form as follows:
\bae{
V(\phi,\psi)=\frac{1}{2}m_\phi^2\phi^2 + \frac{1}{2}m_\psi^2\psi^2.
}
We set the mass parameters as $(m_\phi,m_\psi)=\left(0.1, \frac{0.1}{9}\right)$, and the initial value $\bfX_\ini$ as $\phi_\ini=4.50$, $\psi_\ini=6.91$, $\pi_{\phi,\ini}=-0.0794$, and $\pi_{\psi,\ini}=-0.00146$, which is the value of $(\bfphi_{\rm ini},\boldsymbol{\pi}_{\rm ini})$ at backward e-fold $18$ on the noiseless slow-roll trajectory.
In this setting, $\phi$ rolls down to its potential minimum $\phi=0$ before $\psi$, and after that the model is effectively the single-field one with $\psi$ only.
We define the \ac{EOI} surface by $\psi=1.00086$, the value of $\psi$ when the slow-roll parameter $\epsilon_H \coloneqq -\dot{H}/H^2$ becomes 1 on the slow-roll trajectory.

In Figure~\ref{fig:doublechaotic}, we plot $\calB(\NL,\NS)$ with $\NL-\NS=2$ calculated by Algorithm~\ref{alg:main}.
When the potential is separable, the semi-analytical, non-stochastic method to calculate $\calB_\zeta$ developed in Ref.~\cite{Tada_2017} can be used, so we also show the result of this method.
For the forward formulation, the \ac{CR}-based result is shown too.
As shown in the figure, the results of these two methods roughly coincide, which indicates that the proposed method also works in the multi-field inflation model that yields the bispectrum deviating from the \ac{CR}.

\begin{figure}
\centering

\begin{subfigure}{0.49\linewidth}
\centering
\includegraphics[width=\linewidth]{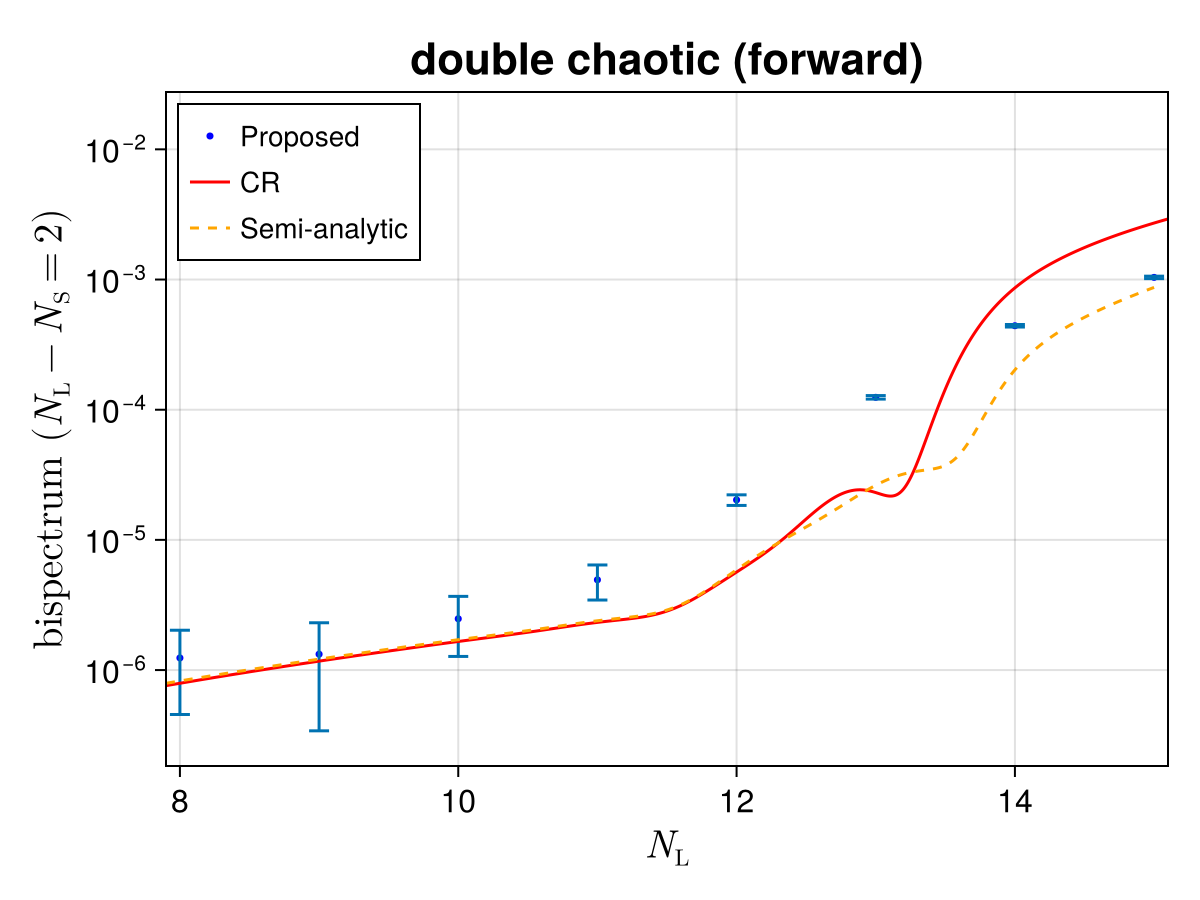}
\caption{Forward formulation}
\end{subfigure}
\hfill
\begin{subfigure}{0.49\linewidth}
\centering
\includegraphics[width=\linewidth]{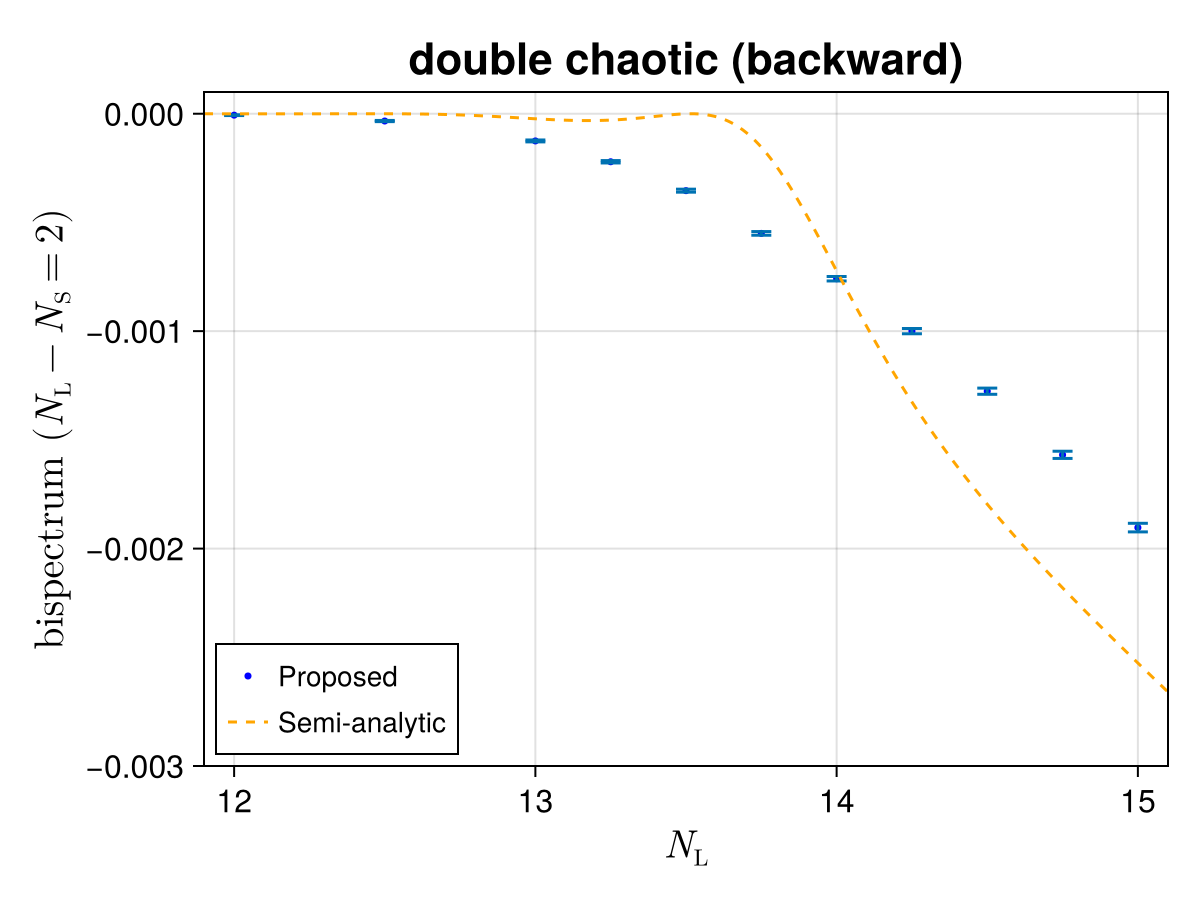}
\caption{Backward formulation}
\end{subfigure}

\caption{Same as Figure~\ref{fig:doublechaotic} except for masses set to $(m_\phi,m_\psi)=\left(0.4, \frac{0.4}{9}\right)$ and the initial momenta set accordingly.}
\label{fig:doublechaotic_largem}
\end{figure}

We also calculate the bispectrum with larger masses $(m_\phi,m_\psi)=\left(0.4, \frac{0.4}{9}\right)$ and the initial value $\phi_\ini=4.50$, $\psi_\ini=6.91$, $\pi_{\phi,\ini}=-0.318$, and $\pi_{\psi,\ini}=-0.00585$.
Figure~\ref{fig:doublechaotic_largem} shows the result.
Now, the results of our method and the method in Ref.~\cite{Tada_2017} show similar tendencies but deviate by an $O(1)$ factor.
In the current setting, the inflation energy scale is larger than in the previous setting, and so is the field fluctuation amplitude, which means a larger stochastic effect.
It implies the necessity of our stochastic formulation beyond the non-stochastic one~\cite{Tada_2017} in such a case.

\subsection{Hybrid inflation}\label{sec: hybrid}

Finally, we consider the hybrid inflation~\cite{Linde:1993cn} involving two fields $\bfphi=(\phi,\psi)$, which are usually called the inflaton and the waterfall field, respectively. 
We adopt the following series-expansion form of the potential for a general discussion:\footnote{$V$ should be considered as an effective potential that is valid in the region of $(\phi,\psi)$ of current interest, with $V_0$ and $V_1$ regarded as the zeroth-order potential and the correction terms, respectively. We assume that the true potential, which is exact in the entire $(\phi,\psi)$ space, is bounded from below, although $V$ in Eq.~\eqref{eq:VHyb} is not.}
\beae{\label{eq:VHyb}
V(\phi,\psi) & = V_0(\phi,\psi) + V_1(\phi,\psi), \\
V_0(\phi,\psi) & = \Lambda^4 \left[\left(1-\left(\frac{\psi}{M}\right)^2\right)^2+2\left(\frac{\phi\psi}{\phi_\uc M}\right)^2\right], \\
V_1(\phi,\psi) & = \Lambda^4 \left[\frac{\phi-\phi_\uc}{\mu_1}-\left(\frac{\phi-\phi_\uc}{\mu_2}\right)^2\right].
}
We set the parameters as
\bae{
M=\frac{\phi_{\rm c}}{\sqrt{2}}=10^{16}\,\si{GeV} \qc \mu_1=\frac{100}{M^2 \phi_{\rm c}} \qc \mu_2=10 \qc \Lambda=\SI{5.4e15}{GeV}\times M\phi_{\rm c}^{1/2},
}
the initial value as
\bae{
\phi_\ini=\phi_{\rm c}+\frac{15}{\mu_1} \qc \psi_\ini=\sqrt{\frac{5\Lambda^4}{24\sqrt{2\pi^3}}},\pi_{\phi,\ini}=\pi_{\psi,\ini}=0,
}
and the \ac{EOI} surface by $\eta_\psi \coloneqq \frac{\partial_\psi^2 V}{V}=-2$, which are the same as Ref.~\cite{miyamoto2025calculating}.
As is typical for this model, in the current setting, $\phi$ rolls down from $\phi_\ini$ to $\phi_{\rm c}$ while $\psi$ remains around $\psi=0$, the potential minimum in the $\psi$ direction.
Then, when $\phi$ reaches $\phi_{\rm c}$, which occurs at backward e-folds around 18 in the current setting, the waterfall transition occurs: $\psi=0$ changes from a minimum to a maximum, causing the fields to fall toward either of the global minima $(0,\pm M)$. 
Furthermore, as is observed in Ref.~\cite{miyamoto2025calculating}, the above parameters make the potential around the waterfall point very flat, and thus the fields' dynamics around the point become diffusion-dominated.
Therefore, the curvature perturbation of a scale that exits the Hubble horizon around the waterfall is largely enhanced, and the power spectrum $\calP_\zeta$ has a peak at that scale.
Thus, we are naturally motivated to investigate the bispectrum around that scale as well.
Besides, since the field values proceed along a path as a curved trajectory with fluctuations, we expect the bispectrum that deviates from the \ac{CR} in the forward formulation and the nonzero bispectrum in the backward formulation.

\begin{figure}
\centering

\begin{subfigure}{0.49\linewidth}
\centering
\includegraphics[width=\linewidth]{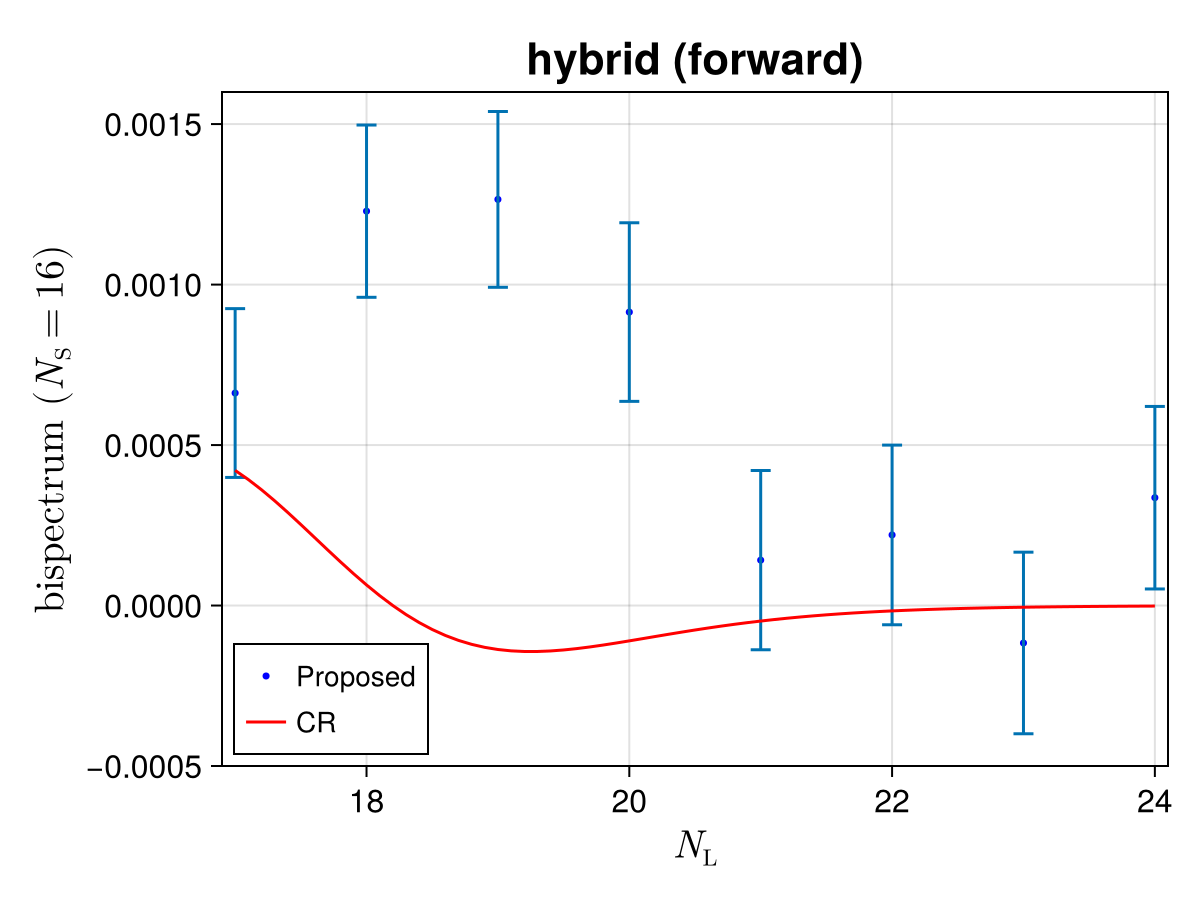}
\caption{Forward formulation}
\end{subfigure}
\hfill
\begin{subfigure}{0.49\linewidth}
\centering
\includegraphics[width=\linewidth]{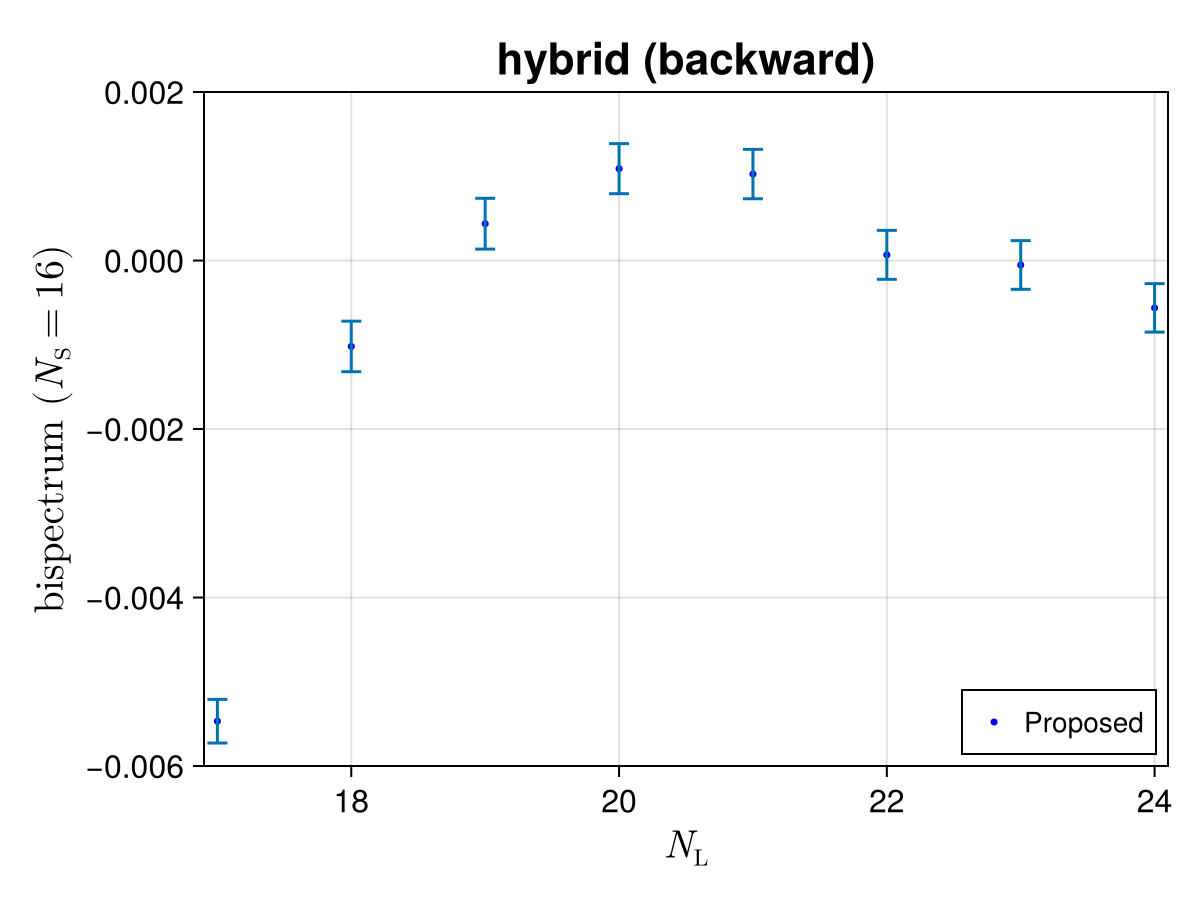}
\caption{Backward formulation}
\end{subfigure}

\caption{Bispectrum in the hybrid inflation in the forward and backward formulation calculated by Algorithm \ref{alg:main} (blue points with error bars). In the forward formulation case, we also show the \ac{CR}-based bispectrum as the red line.}
\label{fig:hybrid}
\end{figure}

We calculate $\calB(\NL,\NS)$ with $\NS$ fixed to 16, the value of $N_\bk$ around which $\calP_\zeta(N_\bk)$ has a peak (see the result in Ref.~\cite{miyamoto2025calculating}), by Algorithm~\ref{alg:main} and plot it in Figure~\ref{fig:hybrid}.
Unlike double-field chaotic inflation, to the best of our knowledge, there is no known semi-analytical method applicable to hybrid inflation.
For comparison, we include the \ac{CR}-based bispectrum in the figure in the forward formulation.
As expected, we obtain the bispectrum deviating from the \ac{CR} in the forward formulation and the nonzero bispectrum in the backward formulation.
Furthermore, the bispectrum converges to 0 for larger $\NL$ with $\NS$ fixed, which is natural in light of the intuition that perturbations of scales separated by many orders of magnitude have a small correlation.

\section{Discussion and Conclusions} \label{sec:sum}

In this paper, we proposed a Monte Carlo-based method to calculate the squeezed bispectrum of the curvature perturbation in the stochastic-$\delta\calN$ formalism. By extending our previous algorithm for the power spectrum~\cite{miyamoto2025calculating}, we avoided the nested path generation and realised a practical calculation of the bispectrum.
Our algorithm can switch between the forward and backward formulations: the former directly relates to the commonly used bispectrum defined in the comoving coordinate, while the latter removes the \emph{gauge-dependent} part and indicates the \emph{physical} correlation between the long- and short-wavelength modes~\cite{Tada_2017}. We demonstrated our algorithm in single-field chaotic inflation (Sec.~\ref{sec: single chaotic}), double-field chaotic inflation (Sec.~\ref{sec: double chaotic}), and hybrid inflation (Sec.~\ref{sec: hybrid}). The first example validated our algorithm, the second one indicates a stochastic correction on the bispectrum, and in the last example, we had success in a quantitative calculation of the squeezed bispectum for the first time in the mild-waterfall variant of hybrid inflation.

The right panel of Figure~\ref{fig:hybrid} exhibits that the peak scale curvature perturbation $\zeta_\rmS$ is negatively correlated to slightly longer modes ($0<N_\rmL-N_\rmS\lesssim2$) but positively correlated to mildly longer modes ($2\lesssim N_\rmL-N_\rmS\lesssim6$), and has zero-consistent correlation to longer enough modes ($6\lesssim N_\rmL-N_\rmS$). Since the enhanced curvature perturbation at the peak scale is a subject for the formation of peculiar astrophysical objects such as \acp{PBH}, the effect of these non-Gaussianities on the formation should also be investigated (see, e.g., Ref.~\cite{Pi:2024lsu} for a recent review).
The positive non-Gaussianity enhances the \ac{PBH} abundance, while the negative one reduces it.
Furthermore, the long-short correlation causes a spatial modulation of the \ac{PBH} production, in the same mechanism as the \emph{scale-dependent bias of halos} (see, e.g., Ref.~\cite{Tada:2015noa}). The spatial modulation, or the \emph{\ac{PBH} clustering}, can affect the merger rate of the \ac{PBH} binaries, and hence it is important in the gravitational wave perspectives of \acp{PBH}.
We leave the detailed analysis in hybrid inflation for future work.

Improving the proposed algorithm is also an important future work.
In path generation in our numerical demonstrations, to suppress the error sufficiently, we set $N_{\rm samp}$, the number of samples, to the large value of $10^6$ and the time step size $\Delta N$ to the small value of $0.01$.
This results in a path-generation time of several hours for obtaining $\hat{\mathcal{B}}(N_\rmL,N_\rmS)$ at a single value of $(N_\rmL,N_\rmS)$, so each graph in Figures \ref{fig:chaotic} to \ref{fig:hybrid} contains points for only several values of $N_\rmL$ with either $N_\rmL-N_\rmS$ or $N_\rmS$ fixed. 
The path generation may be sped up by using advanced Monte Carlo techniques, such as the multilevel Monte Carlo method~\cite{giles2008multilevel}, and/or computing resources such as GPUs.
Speeding up the algorithm enables us to estimate the bispectrum for a larger number of values of $(N_\rmL,N_\rmS)$ with higher accuracy, leading to more detailed discussions of cosmological consequences including \ac{PBH} production.

\section*{Acknowledgements}

KM is supported by MEXT Quantum Leap Flagship Program (MEXT Q-LEAP) Grant No. JPMXS0120319794, and JST COI-NEXT Program Grant No. JPMJPF2014.
YT is supported by JSPS KAKENHI Grant No.~JP24K07047.

\appendix

\bibliographystyle{JHEP}
\bibliography{reference}

\end{document}